\documentclass[sigconf]{acmart}
\AtBeginDocument{%
  }

\usepackage{amsmath}
\usepackage[utf8]{inputenc} 
\usepackage[T1]{fontenc}    
\usepackage{hyperref}       
\usepackage{url}            
\usepackage{booktabs}       
\usepackage{amsfonts}       
\usepackage{nicefrac}       
\usepackage{microtype}      
\usepackage{multirow}
\usepackage{subfigure,comment,adjustbox}
\usepackage{makecell}

\usepackage{subcaption}
\usepackage{graphicx}
\usepackage{xspace}
\usepackage{graphicx}
\definecolor{myblue}{rgb}{0.631, 0.682, 0.765}
\definecolor{myorange}{rgb}{0.749, 0.376, 0.0}
\usepackage{listings}
\usepackage{subcaption}

\usepackage[most]{tcolorbox}

\copyrightyear{2025}
\acmYear{2025}
\setcopyright{acmlicensed}
\acmConference[CIKM '25]{Proceedings of the 34th ACM International Conference on Information and Knowledge Management}{November 10--14, 2025}{Seoul, Republic of Korea}
\acmBooktitle{Proceedings of the 34th ACM International Conference on Information and Knowledge Management (CIKM '25), November 10--14, 2025, Seoul, Republic of Korea}
\acmISBN{979-8-4007-2040-6/2025/11}
\acmDOI{10.1145/XXXXXXX.XXXXXXX}  

\begin{document}

\title{RankArena: A Unified Platform for Evaluating Retrieval, Reranking and RAG with Human and LLM Feedback
}

\author{Abdelrahman Abdallah}
\affiliation{%
  \institution{University of Innsbruck}
  \city{Innsbruck}
  \state{Tyrol}
  \country{Austria}
  }

\author{Mahmoud Abdalla}
\affiliation{%
  \institution{ Chungbuk National University}
  \city{Cheongju-si}
  \state{Cheongju}
  \country{Republic of Korea}
  }

\author{Bhawna Piryani}
\affiliation{%
  \institution{University of Innsbruck}
  \city{Innsbruck}
  \state{Tyrol}
  \country{Austria}
  }

\author{Jamshid Mozafari}
\affiliation{%
  \institution{University of Innsbruck}
 \city{Innsbruck}
  \state{Tyrol}
  \country{Austria}
  }

\author{Mohammed Ali}
\affiliation{%
  \institution{University of Innsbruck}
  \city{Innsbruck}
  \state{Tyrol}
  \country{Austria}
  }

\author{Adam Jatowt}
\affiliation{%
  \institution{University of Innsbruck}
  \city{Innsbruck}
 \state{Tyrol}
  \country{Austria}
  }

\renewcommand{\shortauthors}{Abdallah et al.}

\begin{abstract}
Evaluating the quality of retrieval-augmented generation (RAG) and document reranking systems remains challenging due to the lack of scalable, user-centric, and multi-perspective evaluation tools. We introduce RankArena, a unified platform for comparing and analysing the performance of retrieval pipelines, rerankers, and RAG systems using structured human and LLM-based feedback as well as for collecting such feedback. RankArena supports multiple evaluation modes: direct reranking visualisation, blind pairwise comparisons with human or LLM voting, supervised manual document annotation, and end-to-end RAG answer quality assessment. It captures fine-grained relevance feedback through both pairwise preferences and full-list annotations, along with auxiliary metadata such as movement metrics, annotation time, and quality ratings. The platform also integrates LLM-as-a-judge evaluation, enabling comparison between model-generated rankings and human ground truth annotations. All interactions are stored as structured evaluation datasets that can be used to train rerankers, reward models, judgment agents, or retrieval strategy selectors. Our platform is publicly available at \url{https://rankarena.ngrok.io/}, and the Demo video is provided\footnote{\url{https://youtu.be/jIYAP4PaSSI}}.
\end{abstract}




\maketitle
\vspace{3mm}
\section{Introduction}

Recent advancements in large language models (LLMs) and retrieval-augmented generation (RAG)~\cite{lewis2020retrieval,gao2023retrieval,abdallah2025tempretriever,abdallah2025retrieval,abdallah2023generator,abdallah2023exploring,piryani2024detecting} systems have significantly expanded the scope of natural language understanding and generation. These systems combine information retrieval with powerful generative models, enabling applications such as open-domain question answering~\cite{han2024rag,yang2015wikiqa,wallat2025study,mozafari2024exploring,gruber2024complextempqa}, summarization~\cite{edge2024local}, and chat assistants~\cite{raja2024rag}. However, as models and pipelines become more complex, evaluating their performance in alignment with human preferences remains a persistent challenge. Conventional benchmarks~\cite{thakur2021beir,voorhees2001trec} often relying on static datasets and predefined metrics fail to capture user-centric notions of quality such as \textit{document relevance}, \textit{answer usefulness}, or \textit{faithfulness in generation}.
\begin{figure*}
    \centering
    \includegraphics[width=0.7\linewidth]{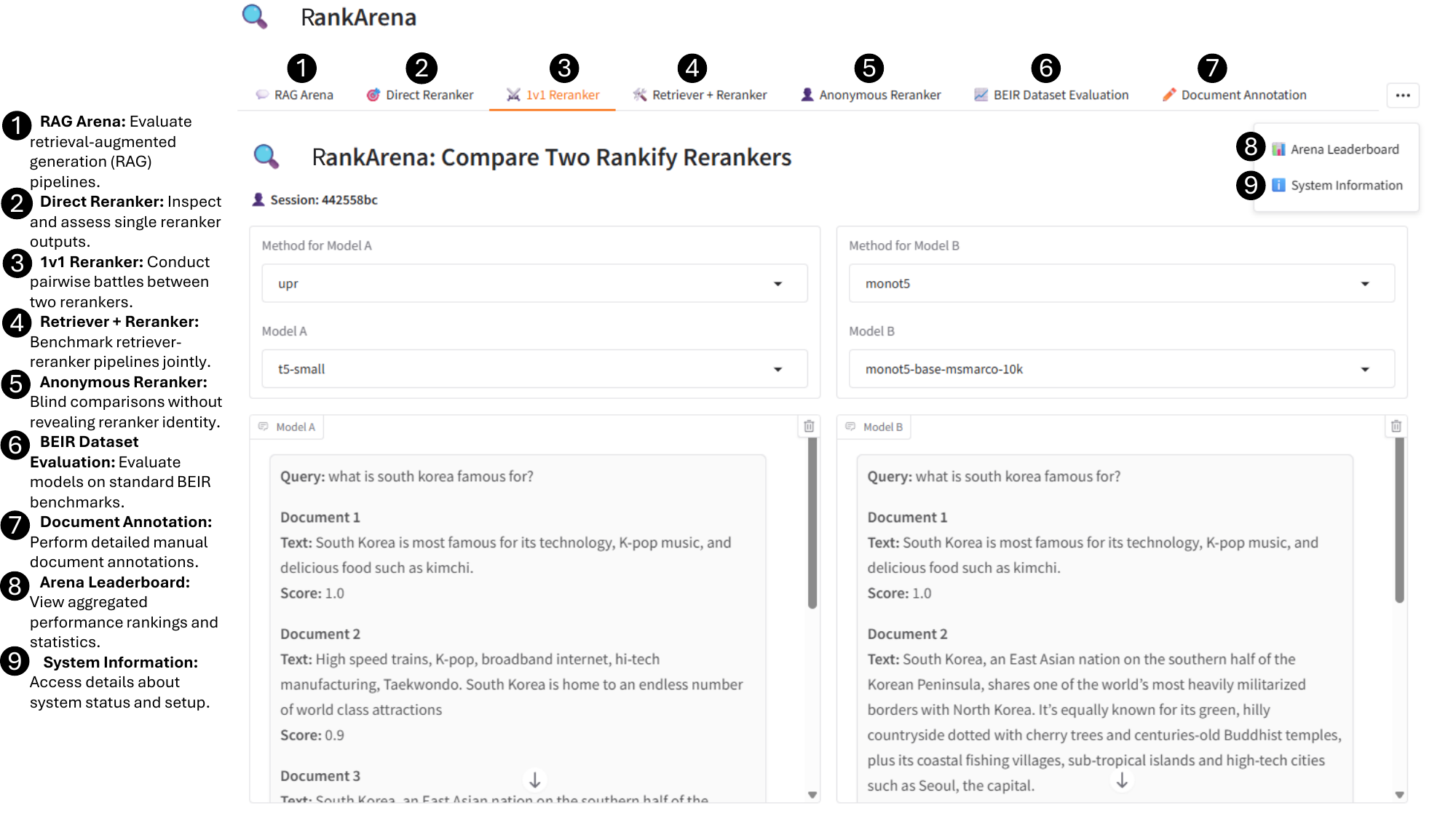}
    \caption{Screenshot of the RankArena interface with added explanations of each tab (1-9) showcasing the system's main evaluation modes and tools.}
    \label{fig:screenshot}
\end{figure*}
To address this gap, we introduce \textbf{RankArena}, a unified, open-source evaluation and annotation platform that supports multi-faceted assessment of \textit{retrieval}, \textit{reranking}, and \textit{generation quality}. Inspired by recent efforts such as MT-Bench~\cite{bai2024mt} and Chatbot Arena~\cite{chiang2024chatbot}, our platform extends the evaluation paradigm beyond model-centric metrics by integrating \textit{human-in-the-loop} and \textit{LLM-as-a-judge} feedback~\cite{zheng2023judging,gu2024survey}. Users interact with multiple reranking and RAG pipelines across various tasks, cast preferences in blind pairwise comparisons, annotate ranked document lists, and evaluate generated answers, all of which are structured into reusable datasets.

RankArena features five complementary evaluation modes:
\begin{enumerate}
    \item \textbf{Reranker Comparison:} Direct and blind pairwise battles between ranked document lists.
    \item \textbf{Manual Annotation:} Supervised full-list ranking with quality labels, tracking annotation effort and movement metrics.
    \item \textbf{LLM Judgment:} GPT-based voting on document orderings or RAG answers.
    \item \textbf{Comprehensive Reranker Leaderboard:} Evaluation and comparison of multiple rerankers across diverse retrieval tasks, using shared datasets and unified metrics.
    \item \textbf{RAG Output Evaluation:} Qualitative and preference-based evaluation of generated answers from end-to-end RAG pipelines.
    \item \textbf{Dataset Collection for Reward and Judge Model Training:} Systematically collect aligned human and LLM preference data, creating reusable datasets to train reward models for retrieval and to build or fine-tune LLM judges capable of comparing retrieval or RAG systems effectively.

\end{enumerate}

Unlike previous works that focus solely on chatbot alignment or answer preference~\cite{chiang2024chatbot,luo2024wizardarena}, RankArena supports \textit{retrieval-specific evaluation signals}, making it uniquely suited to train and benchmark rerankers, reward models, and retrieval agents. Our platform captures diverse signals, including user preferences, annotation metadata, rank movements, generation ratings, and LLM feedback, all stored in a structured, extensible dataset format. These datasets can be used for \textit{training new models}, analyzing \textit{alignment gaps} between humans and LLMs, or developing \textit{retrieval strategies adaptive to user feedback}.



\begin{figure}[t]
\centering
\includegraphics[width=0.7\linewidth]{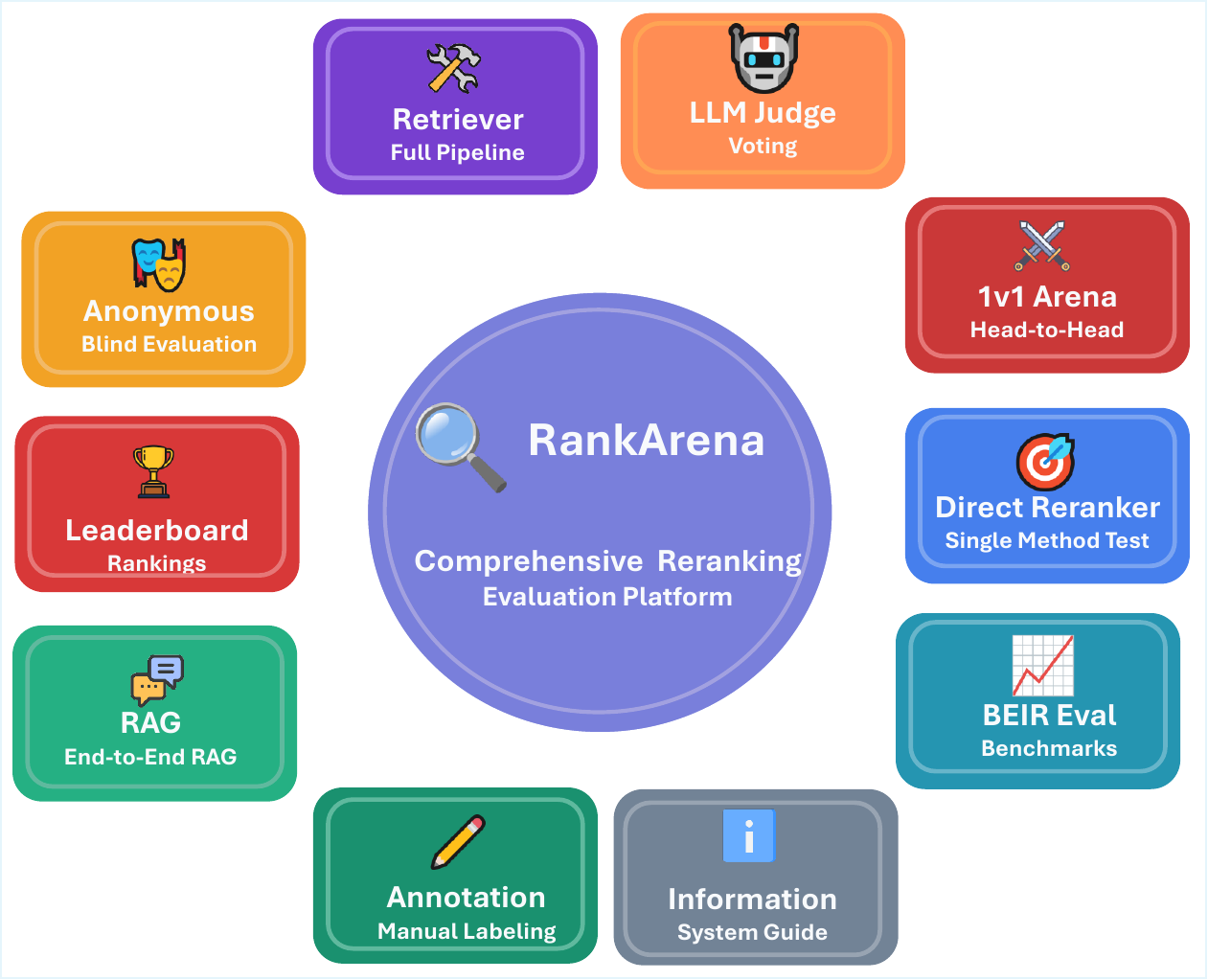}
\caption{The overview of RankArena highlighting its main evaluation modules.}
\label{fig:overview}
\end{figure}
\section{RankArena Functionality}

RankArena fills this gap by providing a unified platform for collecting pairwise preferences, listwise annotations, and LLM-as-a-judge feedback at scale. This system is important because it enables:
(i) creation of reusable datasets for training supervised rerankers and reward models;
(ii) study of human-LLM agreement in ranking tasks;
(iii) adaptive benchmarking that evolves with real user queries and feedback;
and (iv) reproducible, open evaluations combining human and automated signals. 

The overview of our system is illustrated in Figure~\ref{fig:screenshot} and ~\ref{fig:overview}, which highlights the various modules working in concert to support rich evaluation scenarios. Our platform facilitates blind head-to-head comparisons, direct inspection of retrieval outputs, large-scale annotation, and statistical aggregation of pairwise preferences into global model rankings.

\paragraph{Arena Battles and Preference Collection:}

At the heart of RankArena lies the \textbf{1v1 Arena}, a head-to-head evaluation interface where two rerankers, retrieval systems, or RAG pipelines are presented in a blind and randomized manner. Either human or LLM-based judges interact with the system by comparing the outputs of these two systems on the same query and voting for the preferred result. This design simplifies the cognitive load on evaluators compared to absolute scoring, as it only requires relative preference judgments.  We support over 24 reranking methods~\cite{rashid2024ecorank,llm2vsc,jiang2023llmblender,chen2024early,laitz2024inranker,formal2022distillation,khattab2020colbert,Damodaran_FlashRank_Lightest_and_2023,pradeep2023rankzephyr,pradeep2023rankvicuna,tamber2023scalingdownlittingup,tamber2023scalingdownlittingup,zhuang2022rankt5finetuningt5text,nogueira2019multistagedocumentrankingbert,abdallah2025asrank,abdallah2024dynrank} spanning pairwise, listwise, and pointwise approaches, with a total of 84 models available for evaluation in these head-to-head battles.

\paragraph{LLM-as-a-Judge Evaluation:}

To complement human annotations, RankArena integrates an \textbf{LLM judge} module, where large language models (e.g., GPT-4) provide automated pairwise preferences between retrieval or RAG outputs. The LLM-as-a-judge mechanism enables rapid scaling of evaluations and offers a cost-effective alternative to human voting. We adopt a structured prompting strategy that asks LLMs to consider relevance, ranking order quality, and overall usefulness in answering the query. The prompt clearly instructs the LLM to review two reranked document lists for a given query and respond strictly within a predefined format, specifying the winning reranker and providing concise reasoning. This approach ensures consistency, interpretability, and comparability of LLM judgments across different evaluation tasks. 
\paragraph{End-to-End RAG:}

RankArena also supports the evaluation of \textbf{full RAG pipelines}, where annotators or LLM judges assess generated answers along with their supporting document lists. This integrated evaluation allows for measuring not only document ranking quality but also how well retrieved evidence supports final answer generation. Such evaluation is essential for understanding system performance in complex, multi-stage pipelines. 
We support five different retrievers (DPR~\cite{karpukhin2020dense}, Colbert~\cite{khattab2020colbert}, Contreiver~\cite{izacard2021contriever}, BGE~\cite{chen2024bge}, BM25~\cite{robertson2009probabilistic}) operating over two indexing corpora—\textbf{Wikipedia} and \textbf{MS MARCO} as well as online retrievers for live document retrieval, providing flexibility across static and dynamic knowledge sources. 

\begin{figure}[t]
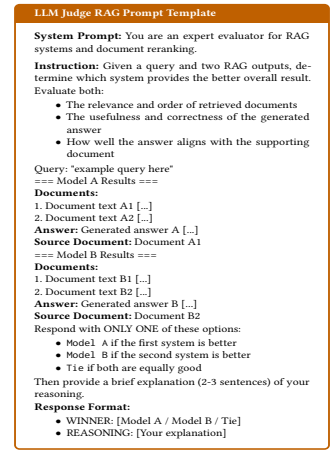

\centering
\resizebox{0.5\linewidth}{0.7\linewidth}{%
\begin{tcolorbox}[colback=blue!0!white, colframe=orange!75!black, title=LLM Judge RAG Prompt Template, fonttitle=\bfseries]
\textbf{System Prompt:} You are an expert evaluator for RAG systems and document reranking.

\vspace{0.5em}
\textbf{Instruction:} Given a query and two RAG outputs, determine which system provides the better overall result. Evaluate both:
\begin{itemize}
    \item The relevance and order of retrieved documents
    \item The usefulness and correctness of the generated answer
    \item How well the answer aligns with the supporting document
\end{itemize}

\noindent Query: "example query here"

\noindent === Model A Results ===\\
\textbf{Documents:}\\
1. Document text A1 [...]\\
2. Document text A2 [...]\\
\textbf{Answer:} Generated answer A [...]\\
\textbf{Source Document:} Document A1

\noindent === Model B Results ===\\
\textbf{Documents:}\\
1. Document text B1 [...]\\
2. Document text B2 [...]\\
\textbf{Answer:} Generated answer B [...]\\
\textbf{Source Document:} Document B2

\noindent Respond with ONLY ONE of these options:
\begin{itemize}
    \item \texttt{Model A} if the first system is better
    \item \texttt{Model B} if the second system is better
    \item \texttt{Tie} if both are equally good
\end{itemize}

\noindent Then provide a brief explanation (2-3 sentences) of your reasoning.

\noindent \textbf{Response Format:}
\begin{itemize}
    \item WINNER: [Model A / Model B / Tie]
    \item REASONING: [Your explanation]
\end{itemize}
\end{tcolorbox}
}
\caption{Illustration of our LLM Judge RAG prompt template. The LLM evaluates RAG results by comparing retrieved documents, generated answers, and evidence alignment.}
\label{fig:llm-judge-rag-prompt}
\end{figure}

For RAG evaluations, we extend our \textbf{LLM-as-a-Judge} module to assess both the generated answers and their associated supporting documents from two different rerankers. The LLM judge considers not only the relevance and order of retrieved contexts but also the quality of the final answers and their alignment with the retrieved evidence. This method enables scalable, automated assessment of how effectively different rerankers support answer generation in RAG systems. We adopt a structured prompt for RAG evaluations that asks the LLM judge to compare two RAG outputs each consisting of a document list, a generated answer, and an indication of which document the answer was primarily drawn from. The LLM is instructed to evaluate results based on: (1) Relevance of retrieved documents, (2) Quality of document ranking, (3) Usefulness and correctness of the generated answers, and (4) Faithfulness of the answers to the supporting documents.  This structured prompting is illustrated in Figure~\ref{fig:llm-judge-rag-prompt}, enabling consistent and interpretable LLM voting for end-to-end RAG pipelines.

\paragraph{Direct Reranker and Full-List Annotation:}

Beyond head-to-head comparisons, RankArena supports two complementary modes for evaluating single reranker outputs. In the \textbf{direct reranker} mode, users provide both the query and the candidate documents, and the system applies a single reranker to produce a ranked list. This mode allows users to directly examine how a specific reranker orders documents for a given query. In contrast, the \textbf{full-list annotation} mode involves users providing a query (or selecting one), after which the system retrieves documents from online or offline corpora. The user then manually reorders or assigns relevance grades to these documents, creating high-quality listwise supervision data. This annotated data can be used to train supervised rerankers, listwise ranking models, or reward models for preference learning.

\paragraph{Leaderboard Aggregation and Statistical Modeling:}

The preferences collected from head-to-head battles, direct evaluations, and LLM judges are aggregated into a \textbf{comprehensive reranker leaderboard}. We compute win rates \( w_i = \frac{\text{wins}_i}{\text{total votes}_i} \) and transform these into an ELO-style rating:
\[
R_i = 1200 + 32 \cdot (w_i - 0.5) \cdot \min\left( \log(\text{total votes}_i + 1), 5.0 \right)
\]
This formulation reflects both win rate and the confidence in that estimate (via number of votes). Traditional benchmark scores, such as BEIR averages, are calculated as:
\(
\text{BEIR Avg}_i = \frac{1}{N} \sum_{j=1}^N S_{ij}
\)
where \( S_{ij} \) is the score on benchmark \( j \). Models are ranked by descending ELO rating, with BEIR performance as a secondary criterion. Our leaderboard thus integrates user and LLM preference signals with standardized benchmarks, offering a holistic view of model performance.

\section{Results and Analysis}
\paragraph{Experiment Setup}

Our experiments were conducted using the \textbf{RankArena} platform, developed on top of the \texttt{Rankify}~\cite{abdallah2025rankify}, Gradio~\cite{abid2019gradio} and \texttt{PyTorch}~\cite{paszke2019pytorch} framework. The system is deployed on a server equipped with two NVIDIA A40 GPUs and 250 GB of RAM, providing sufficient resources to support continuous 24-hour operation for large-scale evaluation and annotation.

RankArena integrates diverse reranking and retrieval components for comprehensive benchmarking. The system currently includes: (1) \textbf{24 reranking methods} spanning pointwise, pairwise, and listwise approaches, (2) \textbf{84 reranker models} (as some methods apply different models) evaluated across different retrieval tasks, and (3) \textbf{5 retrievers} operating on two corpora: \textbf{Wikipedia} and \textbf{MS MARCO}, along with online retrieval for dynamic evaluation. Below, we report the results and analyses based on the following \textbf{Arena Statistics}: (1) \textbf{Total User Votes:} 102. (2) \textbf{Active Models:} 36 (models that have received human votes). (3) \textbf{Benchmark Models:} 80 (models with BEIR benchmark scores). (4) \textbf{Methods Tested:} 25. (5) \textbf{Average User-LLM Agreement:} 74.2\%

\begin{figure}[t]
\centering
\includegraphics[width=0.8\linewidth]{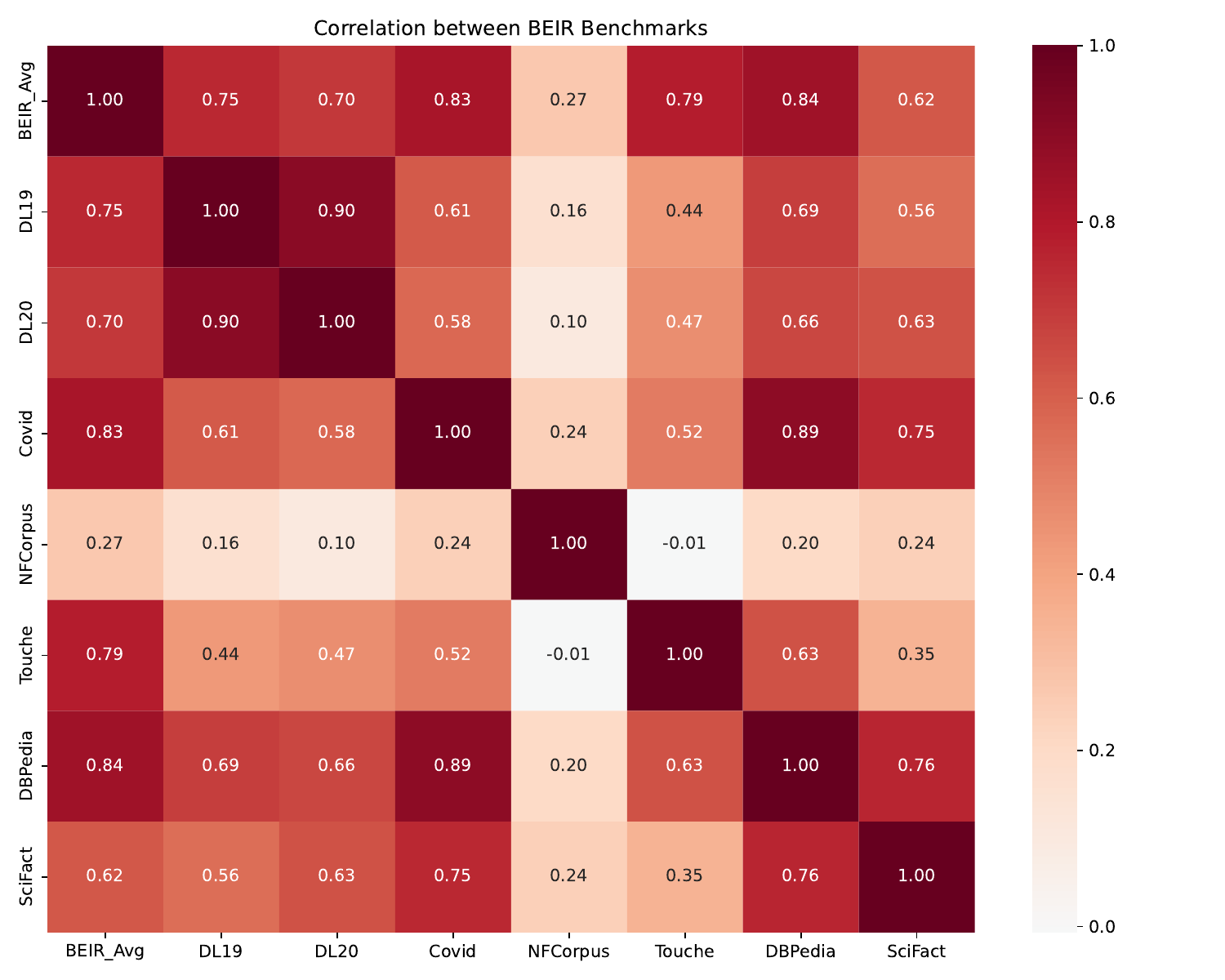}
\caption{Correlation matrix between BEIR Average and individual benchmarks.}
\label{fig:beir_correlation}
\end{figure}

\paragraph{BEIR Benchmark Correlations} We analyze the correlation between different BEIR benchmark~\cite{thakur2021beir} datasets to understand how model performance generalizes across tasks. 
Figure~\ref{fig:beir_correlation} (available also in the leaderboard) illustrates the correlation matrix between BEIR Average and individual benchmarks (DL19, DL20, Covid, NFCorpus, Touche, DBPedia, SciFact).

T       he BEIR Average shows strong correlations with DBPedia (0.84), Covid (0.83), and DL19 (0.75), indicating these datasets contribute substantially to the aggregate BEIR metric. In contrast, NFCorpus displays weak correlation (0.27) with BEIR Average and other benchmarks, suggesting it measures distinct capabilities. Such correlation patterns provide insight into the diversity of tasks within BEIR and the challenges of designing universal rankers.

\paragraph{Models Performance on BEIR}

\begin{figure}[t]
\centering
\includegraphics[width=0.9\linewidth]{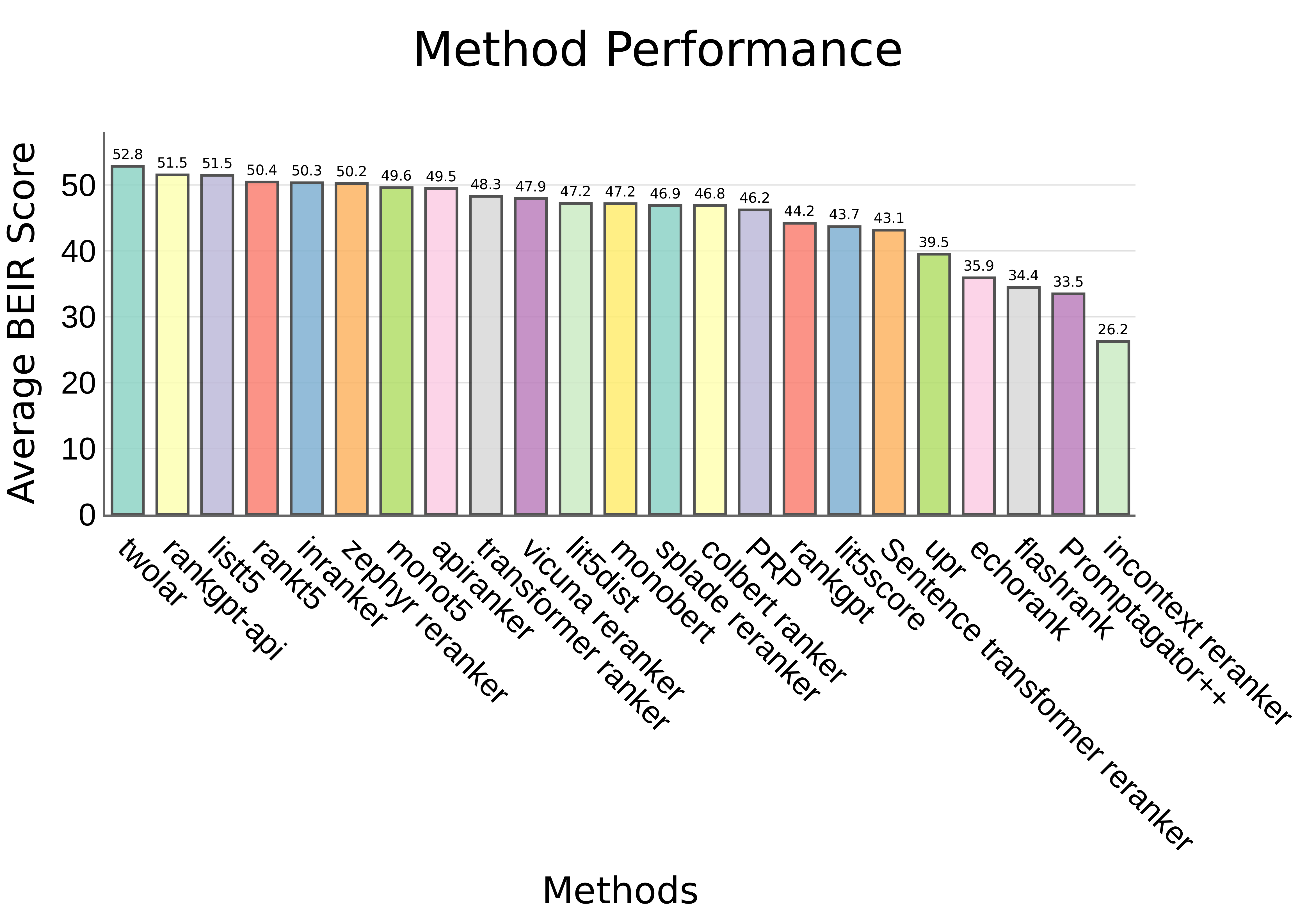}
\caption{Average BEIR performance of different reranking methods.}
\label{fig:beir_performance}
\end{figure}

Figure~\ref{fig:beir_performance} (also displayed in the leaderboard tab) presents the average BEIR scores across evaluated methods.
The highest-performing method, \texttt{twolar}~\cite{baldelli2024twolar}, achieves an average score of 52.8, while other top methods (e.g., \texttt{rankgpt-api}~\cite{sun2023chatgpt}, \texttt{listt5}~\cite{yoon2024listt5}) exhibit scores above 50.
Methods like \texttt{UPR}~\cite{sachan2022improving} and \texttt{incontext reranker}~\cite{chen2024attention} perform considerably worse, reflecting a gap between the strongest and weakest reranking strategies.

\paragraph{Human-LLM Agreement}

\begin{figure}[t]
\centering
\includegraphics[width=0.9\linewidth]{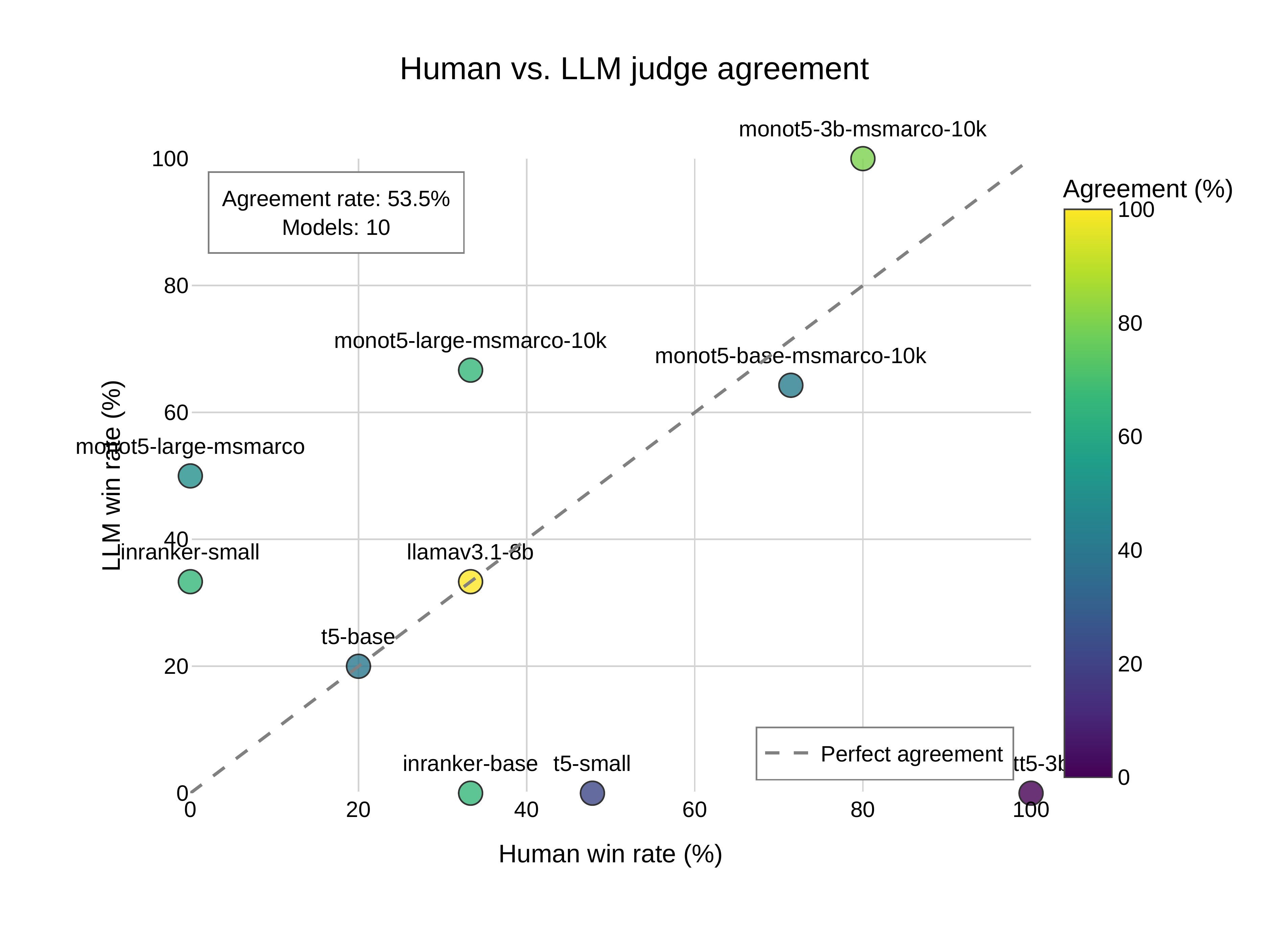}
\caption{Scatter plot of human-LLM agreement across models.}
\label{fig:human_llm_agreement}
\end{figure}

We assess next the alignment between human votes and LLM judge decisions. Figure~\ref{fig:human_llm_agreement} shows a scatter plot of human preference rates versus LLM judge rates across the evaluated models. The results demonstrate a generally strong positive correlation between human and LLM judgments, with most models clustering near the diagonal line. Notable deviations are observed for certain models, indicating areas where human preferences and automated judgments diverge, warranting deeper qualitative analysis.

\begin{figure}
    \centering
    \includegraphics[width=0.8\linewidth]{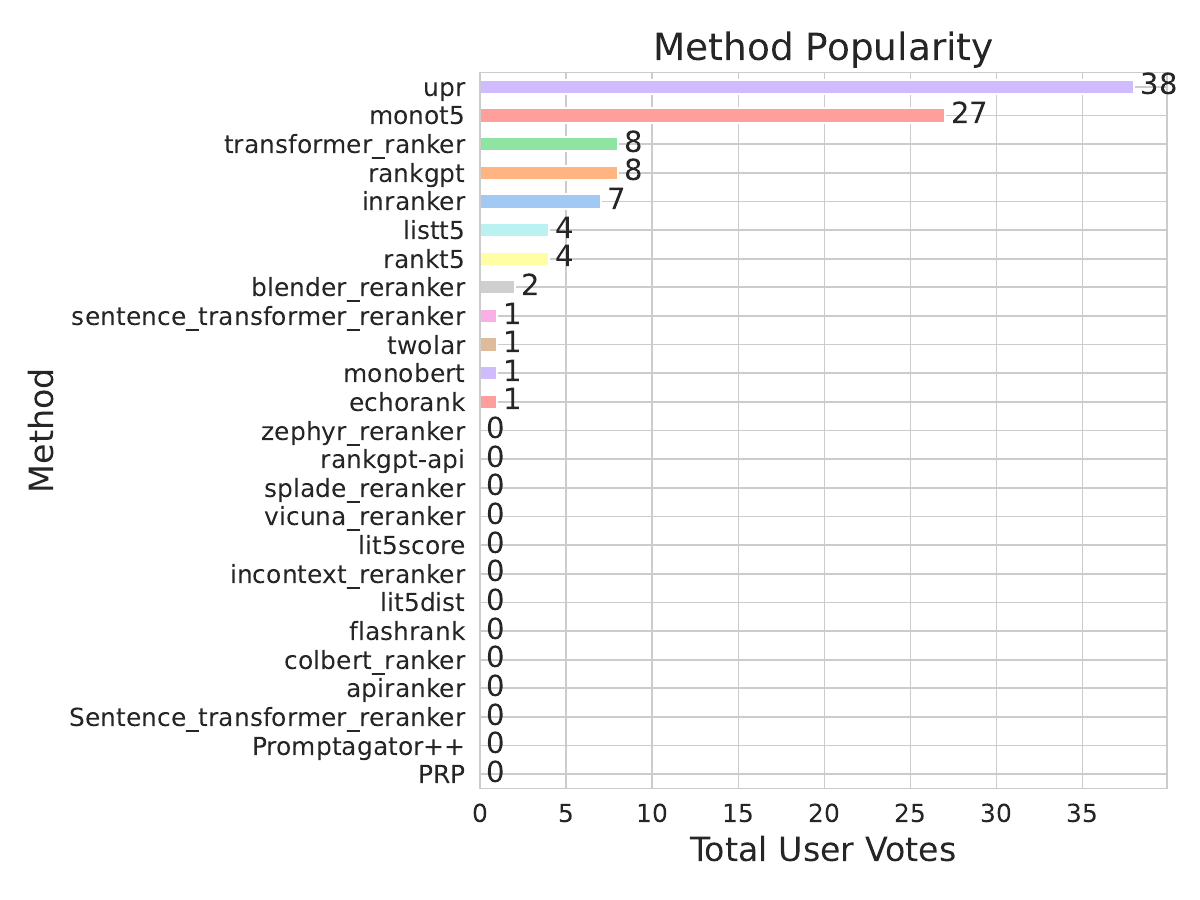}
\caption{Distribution of user votes across different reranking methods in RankArena. The figure shows that a few methods receive }
    \label{fig:enter-label}
\end{figure}

\paragraph{Method Popularity and User Engagement}

Figure~\ref{fig:enter-label} (shown also under the leaderboard tab) illustrates the popularity of different reranking methods in RankArena, measured by the total number of user votes received during head-to-head evaluations. Our data shows a clear skew in user engagement across methods: a small number of rerankers dominate in terms of participation, while many others receive few or no votes. 

\section{Conclusion}
We presented RankArena, a unified platform for evaluating retrieval, reranking, and RAG systems using human and LLM feedback. By combining pairwise preferences, full-list annotations, and automated judgments, RankArena enables the creation of reusable datasets and open benchmarking of retrieval pipelines. 

\section{GenAI Usage Disclosure}
We used OpenAI’s ChatGPT for minor language editing, specifically
to rephrase sentences and correct grammatical errors.

\bibliographystyle{ACM-Reference-Format}
\bibliography{software}


\begin{thebibliography}{48}


\ifx \showCODEN    \undefined \def \showCODEN     #1{\unskip}     \fi
\ifx \showDOI      \undefined \def \showDOI       #1{#1}\fi
\ifx \showISBNx    \undefined \def \showISBNx     #1{\unskip}     \fi
\ifx \showISBNxiii \undefined \def \showISBNxiii  #1{\unskip}     \fi
\ifx \showISSN     \undefined \def \showISSN      #1{\unskip}     \fi
\ifx \showLCCN     \undefined \def \showLCCN      #1{\unskip}     \fi
\ifx \shownote     \undefined \def \shownote      #1{#1}          \fi
\ifx \showarticletitle \undefined \def \showarticletitle #1{#1}   \fi
\ifx \showURL      \undefined \def \showURL       {\relax}        \fi
\providecommand\bibfield[2]{#2}
\providecommand\bibinfo[2]{#2}
\providecommand\natexlab[1]{#1}
\providecommand\showeprint[2][]{arXiv:#2}

\bibitem[Abdallah and Jatowt(2023)]%
        {abdallah2023generator}
\bibfield{author}{\bibinfo{person}{Abdelrahman Abdallah} {and} \bibinfo{person}{Adam Jatowt}.} \bibinfo{year}{2023}\natexlab{}.
\newblock \showarticletitle{Generator-retriever-generator: A novel approach to open-domain question answering}.
\newblock \bibinfo{journal}{\emph{arXiv preprint arXiv:2307.11278}} (\bibinfo{year}{2023}).
\newblock


\bibitem[Abdallah et~al\mbox{.}(2024)]%
        {abdallah2024dynrank}
\bibfield{author}{\bibinfo{person}{Abdelrahman Abdallah}, \bibinfo{person}{Jamshid Mozafari}, \bibinfo{person}{Bhawna Piryani}, \bibinfo{person}{Mohammed~M Abdelgwad}, {and} \bibinfo{person}{Adam Jatowt}.} \bibinfo{year}{2024}\natexlab{}.
\newblock \showarticletitle{DynRank: Improving Passage Retrieval with Dynamic Zero-Shot Prompting Based on Question Classification}.
\newblock \bibinfo{journal}{\emph{arXiv preprint arXiv:2412.00600}} (\bibinfo{year}{2024}).
\newblock


\bibitem[Abdallah et~al\mbox{.}(2025a)]%
        {abdallah2025retrieval}
\bibfield{author}{\bibinfo{person}{Abdelrahman Abdallah}, \bibinfo{person}{Jamshid Mozafari}, \bibinfo{person}{Bhawna Piryani}, \bibinfo{person}{Mohammed Ali}, {and} \bibinfo{person}{Adam Jatowt}.} \bibinfo{year}{2025}\natexlab{a}.
\newblock \showarticletitle{From Retrieval to Generation: Comparing Different Approaches}.
\newblock \bibinfo{journal}{\emph{arXiv preprint arXiv:2502.20245}} (\bibinfo{year}{2025}).
\newblock


\bibitem[Abdallah et~al\mbox{.}(2025b)]%
        {abdallah2025asrank}
\bibfield{author}{\bibinfo{person}{Abdelrahman Abdallah}, \bibinfo{person}{Jamshid Mozafari}, \bibinfo{person}{Bhawna Piryani}, {and} \bibinfo{person}{Adam Jatowt}.} \bibinfo{year}{2025}\natexlab{b}.
\newblock \showarticletitle{Asrank: Zero-shot re-ranking with answer scent for document retrieval}.
\newblock \bibinfo{journal}{\emph{arXiv preprint arXiv:2501.15245}} (\bibinfo{year}{2025}).
\newblock


\bibitem[Abdallah et~al\mbox{.}(2023)]%
        {abdallah2023exploring}
\bibfield{author}{\bibinfo{person}{Abdelrahman Abdallah}, \bibinfo{person}{Bhawna Piryani}, {and} \bibinfo{person}{Adam Jatowt}.} \bibinfo{year}{2023}\natexlab{}.
\newblock \showarticletitle{Exploring the state of the art in legal QA systems}.
\newblock \bibinfo{journal}{\emph{Journal of Big Data}} \bibinfo{volume}{10}, \bibinfo{number}{1} (\bibinfo{year}{2023}), \bibinfo{pages}{127}.
\newblock


\bibitem[Abdallah et~al\mbox{.}(2025c)]%
        {abdallah2025rankify}
\bibfield{author}{\bibinfo{person}{Abdelrahman Abdallah}, \bibinfo{person}{Bhawna Piryani}, \bibinfo{person}{Jamshid Mozafari}, \bibinfo{person}{Mohammed Ali}, {and} \bibinfo{person}{Adam Jatowt}.} \bibinfo{year}{2025}\natexlab{c}.
\newblock \showarticletitle{Rankify: A comprehensive python toolkit for retrieval, re-ranking, and retrieval-augmented generation}.
\newblock \bibinfo{journal}{\emph{arXiv preprint arXiv:2502.02464}} (\bibinfo{year}{2025}).
\newblock


\bibitem[Abdallah et~al\mbox{.}(2025d)]%
        {abdallah2025tempretriever}
\bibfield{author}{\bibinfo{person}{Abdelrahman Abdallah}, \bibinfo{person}{Bhawna Piryani}, \bibinfo{person}{Jonas Wallat}, \bibinfo{person}{Avishek Anand}, {and} \bibinfo{person}{Adam Jatowt}.} \bibinfo{year}{2025}\natexlab{d}.
\newblock \showarticletitle{Tempretriever: Fusion-based temporal dense passage retrieval for time-sensitive questions}.
\newblock \bibinfo{journal}{\emph{arXiv preprint arXiv:2502.21024}} (\bibinfo{year}{2025}).
\newblock


\bibitem[Abid et~al\mbox{.}(2019)]%
        {abid2019gradio}
\bibfield{author}{\bibinfo{person}{Abubakar Abid}, \bibinfo{person}{Ali Abdalla}, \bibinfo{person}{Ali Abid}, \bibinfo{person}{Dawood Khan}, \bibinfo{person}{Abdulrahman Alfozan}, {and} \bibinfo{person}{James Zou}.} \bibinfo{year}{2019}\natexlab{}.
\newblock \showarticletitle{Gradio: Hassle-free sharing and testing of ml models in the wild}.
\newblock \bibinfo{journal}{\emph{arXiv preprint arXiv:1906.02569}} (\bibinfo{year}{2019}).
\newblock


\bibitem[Bai et~al\mbox{.}(2024)]%
        {bai2024mt}
\bibfield{author}{\bibinfo{person}{Ge Bai}, \bibinfo{person}{Jie Liu}, \bibinfo{person}{Xingyuan Bu}, \bibinfo{person}{Yancheng He}, \bibinfo{person}{Jiaheng Liu}, \bibinfo{person}{Zhanhui Zhou}, \bibinfo{person}{Zhuoran Lin}, \bibinfo{person}{Wenbo Su}, \bibinfo{person}{Tiezheng Ge}, \bibinfo{person}{Bo Zheng}, {et~al\mbox{.}}} \bibinfo{year}{2024}\natexlab{}.
\newblock \showarticletitle{Mt-bench-101: A fine-grained benchmark for evaluating large language models in multi-turn dialogues}.
\newblock \bibinfo{journal}{\emph{arXiv preprint arXiv:2402.14762}} (\bibinfo{year}{2024}).
\newblock


\bibitem[Baldelli et~al\mbox{.}(2024)]%
        {baldelli2024twolar}
\bibfield{author}{\bibinfo{person}{Davide Baldelli}, \bibinfo{person}{Junfeng Jiang}, \bibinfo{person}{Akiko Aizawa}, {and} \bibinfo{person}{Paolo Torroni}.} \bibinfo{year}{2024}\natexlab{}.
\newblock \showarticletitle{TWOLAR: a TWO-step LLM-Augmented distillation method for passage Reranking}. In \bibinfo{booktitle}{\emph{European Conference on Information Retrieval}}. Springer, \bibinfo{pages}{470--485}.
\newblock


\bibitem[BehnamGhader et~al\mbox{.}(2024)]%
        {llm2vsc}
\bibfield{author}{\bibinfo{person}{Parishad BehnamGhader}, \bibinfo{person}{Vaibhav Adlakha}, \bibinfo{person}{Marius Mosbach}, \bibinfo{person}{Dzmitry Bahdanau}, \bibinfo{person}{Nicolas Chapados}, {and} \bibinfo{person}{Siva Reddy}.} \bibinfo{year}{2024}\natexlab{}.
\newblock \bibinfo{title}{LLM2Vec: Large Language Models Are Secretly Powerful Text Encoders}.
\newblock
\newblock
\showeprint[arxiv]{2404.05961}~[cs.CL]
\urldef\tempurl%
\url{https://arxiv.org/abs/2404.05961}
\showURL{%
\tempurl}


\bibitem[Chen et~al\mbox{.}(2024c)]%
        {chen2024bge}
\bibfield{author}{\bibinfo{person}{Jianlv Chen}, \bibinfo{person}{Shitao Xiao}, \bibinfo{person}{Peitian Zhang}, \bibinfo{person}{Kun Luo}, \bibinfo{person}{Defu Lian}, {and} \bibinfo{person}{Zheng Liu}.} \bibinfo{year}{2024}\natexlab{c}.
\newblock \showarticletitle{Bge m3-embedding: Multi-lingual, multi-functionality, multi-granularity text embeddings through self-knowledge distillation}.
\newblock \bibinfo{journal}{\emph{arXiv preprint arXiv:2402.03216}} (\bibinfo{year}{2024}).
\newblock


\bibitem[Chen et~al\mbox{.}(2024a)]%
        {chen2024attention}
\bibfield{author}{\bibinfo{person}{Shijie Chen}, \bibinfo{person}{Bernal~Jim{\'e}nez Guti{\'e}rrez}, {and} \bibinfo{person}{Yu Su}.} \bibinfo{year}{2024}\natexlab{a}.
\newblock \showarticletitle{Attention in Large Language Models Yields Efficient Zero-Shot Re-Rankers}.
\newblock \bibinfo{journal}{\emph{arXiv preprint arXiv:2410.02642}} (\bibinfo{year}{2024}).
\newblock


\bibitem[Chen et~al\mbox{.}(2024b)]%
        {chen2024early}
\bibfield{author}{\bibinfo{person}{Zijian Chen}, \bibinfo{person}{Ronak Pradeep}, {and} \bibinfo{person}{Jimmy Lin}.} \bibinfo{year}{2024}\natexlab{b}.
\newblock \bibinfo{title}{An Early FIRST Reproduction and Improvements to Single-Token Decoding for Fast Listwise Reranking}.
\newblock
\newblock
\showeprint[arxiv]{2411.05508}~[cs.IR]
\urldef\tempurl%
\url{https://arxiv.org/abs/2411.05508}
\showURL{%
\tempurl}


\bibitem[Chiang et~al\mbox{.}(2024)]%
        {chiang2024chatbot}
\bibfield{author}{\bibinfo{person}{Wei-Lin Chiang}, \bibinfo{person}{Lianmin Zheng}, \bibinfo{person}{Ying Sheng}, \bibinfo{person}{Anastasios~Nikolas Angelopoulos}, \bibinfo{person}{Tianle Li}, \bibinfo{person}{Dacheng Li}, \bibinfo{person}{Banghua Zhu}, \bibinfo{person}{Hao Zhang}, \bibinfo{person}{Michael Jordan}, \bibinfo{person}{Joseph~E Gonzalez}, {et~al\mbox{.}}} \bibinfo{year}{2024}\natexlab{}.
\newblock \showarticletitle{Chatbot arena: An open platform for evaluating llms by human preference}. In \bibinfo{booktitle}{\emph{Forty-first International Conference on Machine Learning}}.
\newblock


\bibitem[Damodaran(2023)]%
        {Damodaran_FlashRank_Lightest_and_2023}
\bibfield{author}{\bibinfo{person}{Prithiviraj Damodaran}.} \bibinfo{year}{2023}\natexlab{}.
\newblock \bibinfo{booktitle}{\emph{{FlashRank, Lightest and Fastest 2nd Stage Reranker for search pipelines.}}}
\newblock
\urldef\tempurl%
\url{https://doi.org/10.5281/zenodo.10426927}
\showDOI{\tempurl}


\bibitem[Edge et~al\mbox{.}(2024)]%
        {edge2024local}
\bibfield{author}{\bibinfo{person}{Darren Edge}, \bibinfo{person}{Ha Trinh}, \bibinfo{person}{Newman Cheng}, \bibinfo{person}{Joshua Bradley}, \bibinfo{person}{Alex Chao}, \bibinfo{person}{Apurva Mody}, \bibinfo{person}{Steven Truitt}, \bibinfo{person}{Dasha Metropolitansky}, \bibinfo{person}{Robert~Osazuwa Ness}, {and} \bibinfo{person}{Jonathan Larson}.} \bibinfo{year}{2024}\natexlab{}.
\newblock \showarticletitle{From local to global: A graph rag approach to query-focused summarization}.
\newblock \bibinfo{journal}{\emph{arXiv preprint arXiv:2404.16130}} (\bibinfo{year}{2024}).
\newblock


\bibitem[Formal et~al\mbox{.}(2022)]%
        {formal2022distillation}
\bibfield{author}{\bibinfo{person}{Thibault Formal}, \bibinfo{person}{Carlos Lassance}, \bibinfo{person}{Benjamin Piwowarski}, {and} \bibinfo{person}{St{\'e}phane Clinchant}.} \bibinfo{year}{2022}\natexlab{}.
\newblock \showarticletitle{From distillation to hard negative sampling: Making sparse neural ir models more effective}. In \bibinfo{booktitle}{\emph{Proceedings of the 45th international ACM SIGIR conference on research and development in information retrieval}}. \bibinfo{pages}{2353--2359}.
\newblock


\bibitem[Gao et~al\mbox{.}(2023)]%
        {gao2023retrieval}
\bibfield{author}{\bibinfo{person}{Yunfan Gao}, \bibinfo{person}{Yun Xiong}, \bibinfo{person}{Xinyu Gao}, \bibinfo{person}{Kangxiang Jia}, \bibinfo{person}{Jinliu Pan}, \bibinfo{person}{Yuxi Bi}, \bibinfo{person}{Yixin Dai}, \bibinfo{person}{Jiawei Sun}, \bibinfo{person}{Haofen Wang}, {and} \bibinfo{person}{Haofen Wang}.} \bibinfo{year}{2023}\natexlab{}.
\newblock \showarticletitle{Retrieval-augmented generation for large language models: A survey}.
\newblock \bibinfo{journal}{\emph{arXiv preprint arXiv:2312.10997}} \bibinfo{volume}{2}, \bibinfo{number}{1} (\bibinfo{year}{2023}).
\newblock


\bibitem[Gruber et~al\mbox{.}(2024)]%
        {gruber2024complextempqa}
\bibfield{author}{\bibinfo{person}{Raphael Gruber}, \bibinfo{person}{Abdelrahman Abdallah}, \bibinfo{person}{Michael F{\"a}rber}, {and} \bibinfo{person}{Adam Jatowt}.} \bibinfo{year}{2024}\natexlab{}.
\newblock \showarticletitle{Complextempqa: A large-scale dataset for complex temporal question answering}.
\newblock \bibinfo{journal}{\emph{arXiv preprint arXiv:2406.04866}} (\bibinfo{year}{2024}).
\newblock


\bibitem[Gu et~al\mbox{.}(2024)]%
        {gu2024survey}
\bibfield{author}{\bibinfo{person}{Jiawei Gu}, \bibinfo{person}{Xuhui Jiang}, \bibinfo{person}{Zhichao Shi}, \bibinfo{person}{Hexiang Tan}, \bibinfo{person}{Xuehao Zhai}, \bibinfo{person}{Chengjin Xu}, \bibinfo{person}{Wei Li}, \bibinfo{person}{Yinghan Shen}, \bibinfo{person}{Shengjie Ma}, \bibinfo{person}{Honghao Liu}, {et~al\mbox{.}}} \bibinfo{year}{2024}\natexlab{}.
\newblock \showarticletitle{A survey on llm-as-a-judge}.
\newblock \bibinfo{journal}{\emph{arXiv preprint arXiv:2411.15594}} (\bibinfo{year}{2024}).
\newblock


\bibitem[Han et~al\mbox{.}(2024)]%
        {han2024rag}
\bibfield{author}{\bibinfo{person}{Rujun Han}, \bibinfo{person}{Yuhao Zhang}, \bibinfo{person}{Peng Qi}, \bibinfo{person}{Yumo Xu}, \bibinfo{person}{Jenyuan Wang}, \bibinfo{person}{Lan Liu}, \bibinfo{person}{William~Yang Wang}, \bibinfo{person}{Bonan Min}, {and} \bibinfo{person}{Vittorio Castelli}.} \bibinfo{year}{2024}\natexlab{}.
\newblock \showarticletitle{Rag-qa arena: Evaluating domain robustness for long-form retrieval augmented question answering}.
\newblock \bibinfo{journal}{\emph{arXiv preprint arXiv:2407.13998}} (\bibinfo{year}{2024}).
\newblock


\bibitem[Izacard et~al\mbox{.}(2021)]%
        {izacard2021contriever}
\bibfield{author}{\bibinfo{person}{Gautier Izacard}, \bibinfo{person}{Mathilde Caron}, \bibinfo{person}{Lucas Hosseini}, \bibinfo{person}{Sebastian Riedel}, \bibinfo{person}{Piotr Bojanowski}, \bibinfo{person}{Armand Joulin}, {and} \bibinfo{person}{Edouard Grave}.} \bibinfo{year}{2021}\natexlab{}.
\newblock \bibinfo{title}{Unsupervised Dense Information Retrieval with Contrastive Learning}.
\newblock
\newblock
\urldef\tempurl%
\url{https://doi.org/10.48550/ARXIV.2112.09118}
\showDOI{\tempurl}


\bibitem[Jiang et~al\mbox{.}(2023)]%
        {jiang2023llmblender}
\bibfield{author}{\bibinfo{person}{Dongfu Jiang}, \bibinfo{person}{Xiang Ren}, {and} \bibinfo{person}{Bill~Yuchen Lin}.} \bibinfo{year}{2023}\natexlab{}.
\newblock \bibinfo{title}{LLM-Blender: Ensembling Large Language Models with Pairwise Ranking and Generative Fusion}.
\newblock
\newblock
\showeprint[arxiv]{2306.02561}~[cs.CL]
\urldef\tempurl%
\url{https://arxiv.org/abs/2306.02561}
\showURL{%
\tempurl}


\bibitem[Karpukhin et~al\mbox{.}(2020)]%
        {karpukhin2020dense}
\bibfield{author}{\bibinfo{person}{Vladimir Karpukhin}, \bibinfo{person}{Barlas Oguz}, \bibinfo{person}{Sewon Min}, \bibinfo{person}{Patrick~SH Lewis}, \bibinfo{person}{Ledell Wu}, \bibinfo{person}{Sergey Edunov}, \bibinfo{person}{Danqi Chen}, {and} \bibinfo{person}{Wen-tau Yih}.} \bibinfo{year}{2020}\natexlab{}.
\newblock \showarticletitle{Dense Passage Retrieval for Open-Domain Question Answering.}. In \bibinfo{booktitle}{\emph{EMNLP (1)}}. \bibinfo{pages}{6769--6781}.
\newblock


\bibitem[Khattab and Zaharia(2020)]%
        {khattab2020colbert}
\bibfield{author}{\bibinfo{person}{Omar Khattab} {and} \bibinfo{person}{Matei Zaharia}.} \bibinfo{year}{2020}\natexlab{}.
\newblock \showarticletitle{Colbert: Efficient and effective passage search via contextualized late interaction over bert}. In \bibinfo{booktitle}{\emph{Proceedings of the 43rd International ACM SIGIR conference on research and development in Information Retrieval}}. \bibinfo{pages}{39--48}.
\newblock


\bibitem[Laitz et~al\mbox{.}(2024)]%
        {laitz2024inranker}
\bibfield{author}{\bibinfo{person}{Thiago Laitz}, \bibinfo{person}{Konstantinos Papakostas}, \bibinfo{person}{Roberto Lotufo}, {and} \bibinfo{person}{Rodrigo Nogueira}.} \bibinfo{year}{2024}\natexlab{}.
\newblock \bibinfo{title}{InRanker: Distilled Rankers for Zero-shot Information Retrieval}.
\newblock
\newblock
\showeprint[arxiv]{2401.06910}~[cs.IR]
\urldef\tempurl%
\url{https://arxiv.org/abs/2401.06910}
\showURL{%
\tempurl}


\bibitem[Lewis et~al\mbox{.}(2020)]%
        {lewis2020retrieval}
\bibfield{author}{\bibinfo{person}{Patrick Lewis}, \bibinfo{person}{Ethan Perez}, \bibinfo{person}{Aleksandra Piktus}, \bibinfo{person}{Fabio Petroni}, \bibinfo{person}{Vladimir Karpukhin}, \bibinfo{person}{Naman Goyal}, \bibinfo{person}{Heinrich K{\"u}ttler}, \bibinfo{person}{Mike Lewis}, \bibinfo{person}{Wen-tau Yih}, \bibinfo{person}{Tim Rockt{\"a}schel}, {et~al\mbox{.}}} \bibinfo{year}{2020}\natexlab{}.
\newblock \showarticletitle{Retrieval-augmented generation for knowledge-intensive nlp tasks}.
\newblock \bibinfo{journal}{\emph{Advances in neural information processing systems}}  \bibinfo{volume}{33} (\bibinfo{year}{2020}), \bibinfo{pages}{9459--9474}.
\newblock


\bibitem[Luo et~al\mbox{.}(2024)]%
        {luo2024wizardarena}
\bibfield{author}{\bibinfo{person}{Haipeng Luo}, \bibinfo{person}{Qingfeng Sun}, \bibinfo{person}{Can Xu}, \bibinfo{person}{Pu Zhao}, \bibinfo{person}{Qingwei Lin}, \bibinfo{person}{Jian-Guang Lou}, \bibinfo{person}{Shifeng Chen}, \bibinfo{person}{Yansong Tang}, {and} \bibinfo{person}{Weizhu Chen}.} \bibinfo{year}{2024}\natexlab{}.
\newblock \showarticletitle{Wizardarena: Post-training large language models via simulated offline chatbot arena}.
\newblock \bibinfo{journal}{\emph{Advances in Neural Information Processing Systems}}  \bibinfo{volume}{37} (\bibinfo{year}{2024}), \bibinfo{pages}{111544--111570}.
\newblock


\bibitem[Mozafari et~al\mbox{.}(2024)]%
        {mozafari2024exploring}
\bibfield{author}{\bibinfo{person}{Jamshid Mozafari}, \bibinfo{person}{Abdelrahman Abdallah}, \bibinfo{person}{Bhawna Piryani}, {and} \bibinfo{person}{Adam Jatowt}.} \bibinfo{year}{2024}\natexlab{}.
\newblock \showarticletitle{Exploring hint generation approaches in open-domain question answering}.
\newblock \bibinfo{journal}{\emph{arXiv preprint arXiv:2409.16096}} (\bibinfo{year}{2024}).
\newblock


\bibitem[Nogueira et~al\mbox{.}(2019)]%
        {nogueira2019multistagedocumentrankingbert}
\bibfield{author}{\bibinfo{person}{Rodrigo Nogueira}, \bibinfo{person}{Wei Yang}, \bibinfo{person}{Kyunghyun Cho}, {and} \bibinfo{person}{Jimmy Lin}.} \bibinfo{year}{2019}\natexlab{}.
\newblock \bibinfo{title}{Multi-Stage Document Ranking with BERT}.
\newblock
\newblock
\showeprint[arxiv]{1910.14424}~[cs.IR]
\urldef\tempurl%
\url{https://arxiv.org/abs/1910.14424}
\showURL{%
\tempurl}


\bibitem[Paszke(2019)]%
        {paszke2019pytorch}
\bibfield{author}{\bibinfo{person}{A Paszke}.} \bibinfo{year}{2019}\natexlab{}.
\newblock \showarticletitle{Pytorch: An imperative style, high-performance deep learning library}.
\newblock \bibinfo{journal}{\emph{arXiv preprint arXiv:1912.01703}} (\bibinfo{year}{2019}).
\newblock


\bibitem[Piryani et~al\mbox{.}(2024)]%
        {piryani2024detecting}
\bibfield{author}{\bibinfo{person}{Bhawna Piryani}, \bibinfo{person}{Abdelrahman Abdallah}, \bibinfo{person}{Jamshid Mozafari}, {and} \bibinfo{person}{Adam Jatowt}.} \bibinfo{year}{2024}\natexlab{}.
\newblock \showarticletitle{Detecting temporal ambiguity in questions}.
\newblock \bibinfo{journal}{\emph{arXiv preprint arXiv:2409.17046}} (\bibinfo{year}{2024}).
\newblock


\bibitem[Pradeep et~al\mbox{.}(2023a)]%
        {pradeep2023rankvicuna}
\bibfield{author}{\bibinfo{person}{Ronak Pradeep}, \bibinfo{person}{Sahel Sharifymoghaddam}, {and} \bibinfo{person}{Jimmy Lin}.} \bibinfo{year}{2023}\natexlab{a}.
\newblock \bibinfo{title}{RankVicuna: Zero-Shot Listwise Document Reranking with Open-Source Large Language Models}.
\newblock
\newblock
\showeprint[arxiv]{2309.15088}~[cs.IR]
\urldef\tempurl%
\url{https://arxiv.org/abs/2309.15088}
\showURL{%
\tempurl}


\bibitem[Pradeep et~al\mbox{.}(2023b)]%
        {pradeep2023rankzephyr}
\bibfield{author}{\bibinfo{person}{Ronak Pradeep}, \bibinfo{person}{Sahel Sharifymoghaddam}, {and} \bibinfo{person}{Jimmy Lin}.} \bibinfo{year}{2023}\natexlab{b}.
\newblock \bibinfo{title}{RankZephyr: Effective and Robust Zero-Shot Listwise Reranking is a Breeze!}
\newblock
\newblock
\showeprint[arxiv]{2312.02724}~[cs.IR]
\urldef\tempurl%
\url{https://arxiv.org/abs/2312.02724}
\showURL{%
\tempurl}


\bibitem[Raja et~al\mbox{.}(2024)]%
        {raja2024rag}
\bibfield{author}{\bibinfo{person}{Mahimai Raja}, \bibinfo{person}{E Yuvaraajan}, {et~al\mbox{.}}} \bibinfo{year}{2024}\natexlab{}.
\newblock \showarticletitle{A rag-based medical assistant especially for infectious diseases}. In \bibinfo{booktitle}{\emph{2024 International Conference on Inventive Computation Technologies (ICICT)}}. IEEE, \bibinfo{pages}{1128--1133}.
\newblock


\bibitem[Rashid et~al\mbox{.}(2024)]%
        {rashid2024ecorank}
\bibfield{author}{\bibinfo{person}{Muhammad~Shihab Rashid}, \bibinfo{person}{Jannat~Ara Meem}, \bibinfo{person}{Yue Dong}, {and} \bibinfo{person}{Vagelis Hristidis}.} \bibinfo{year}{2024}\natexlab{}.
\newblock \bibinfo{title}{EcoRank: Budget-Constrained Text Re-ranking Using Large Language Models}.
\newblock
\newblock
\showeprint[arxiv]{2402.10866}~[cs.CL]
\urldef\tempurl%
\url{https://arxiv.org/abs/2402.10866}
\showURL{%
\tempurl}


\bibitem[Robertson et~al\mbox{.}(2009)]%
        {robertson2009probabilistic}
\bibfield{author}{\bibinfo{person}{Stephen Robertson}, \bibinfo{person}{Hugo Zaragoza}, {et~al\mbox{.}}} \bibinfo{year}{2009}\natexlab{}.
\newblock \showarticletitle{The probabilistic relevance framework: BM25 and beyond}.
\newblock \bibinfo{journal}{\emph{Foundations and Trends{\textregistered} in Information Retrieval}} \bibinfo{volume}{3}, \bibinfo{number}{4} (\bibinfo{year}{2009}), \bibinfo{pages}{333--389}.
\newblock


\bibitem[Sachan et~al\mbox{.}(2022)]%
        {sachan2022improving}
\bibfield{author}{\bibinfo{person}{Devendra~Singh Sachan}, \bibinfo{person}{Mike Lewis}, \bibinfo{person}{Mandar Joshi}, \bibinfo{person}{Armen Aghajanyan}, \bibinfo{person}{Wen-tau Yih}, \bibinfo{person}{Joelle Pineau}, {and} \bibinfo{person}{Luke Zettlemoyer}.} \bibinfo{year}{2022}\natexlab{}.
\newblock \showarticletitle{Improving Passage Retrieval with Zero-Shot Question Generation}.
\newblock  (\bibinfo{year}{2022}).
\newblock
\urldef\tempurl%
\url{https://arxiv.org/abs/2204.07496}
\showURL{%
\tempurl}


\bibitem[Sun et~al\mbox{.}(2023)]%
        {sun2023chatgpt}
\bibfield{author}{\bibinfo{person}{Weiwei Sun}, \bibinfo{person}{Lingyong Yan}, \bibinfo{person}{Xinyu Ma}, \bibinfo{person}{Shuaiqiang Wang}, \bibinfo{person}{Pengjie Ren}, \bibinfo{person}{Zhumin Chen}, \bibinfo{person}{Dawei Yin}, {and} \bibinfo{person}{Zhaochun Ren}.} \bibinfo{year}{2023}\natexlab{}.
\newblock \showarticletitle{Is ChatGPT good at search? investigating large language models as re-ranking agents}.
\newblock \bibinfo{journal}{\emph{arXiv preprint arXiv:2304.09542}} (\bibinfo{year}{2023}).
\newblock


\bibitem[Tamber et~al\mbox{.}(2023)]%
        {tamber2023scalingdownlittingup}
\bibfield{author}{\bibinfo{person}{Manveer~Singh Tamber}, \bibinfo{person}{Ronak Pradeep}, {and} \bibinfo{person}{Jimmy Lin}.} \bibinfo{year}{2023}\natexlab{}.
\newblock \bibinfo{title}{Scaling Down, LiTting Up: Efficient Zero-Shot Listwise Reranking with Seq2seq Encoder-Decoder Models}.
\newblock
\newblock
\showeprint[arxiv]{2312.16098}~[cs.IR]
\urldef\tempurl%
\url{https://arxiv.org/abs/2312.16098}
\showURL{%
\tempurl}


\bibitem[Thakur et~al\mbox{.}(2021)]%
        {thakur2021beir}
\bibfield{author}{\bibinfo{person}{Nandan Thakur}, \bibinfo{person}{Nils Reimers}, \bibinfo{person}{Andreas R{\"u}ckl{\'e}}, \bibinfo{person}{Abhishek Srivastava}, {and} \bibinfo{person}{Iryna Gurevych}.} \bibinfo{year}{2021}\natexlab{}.
\newblock \showarticletitle{Beir: A heterogenous benchmark for zero-shot evaluation of information retrieval models}.
\newblock \bibinfo{journal}{\emph{arXiv preprint arXiv:2104.08663}} (\bibinfo{year}{2021}).
\newblock


\bibitem[Voorhees(2001)]%
        {voorhees2001trec}
\bibfield{author}{\bibinfo{person}{Ellen~M Voorhees}.} \bibinfo{year}{2001}\natexlab{}.
\newblock \showarticletitle{The TREC question answering track}.
\newblock \bibinfo{journal}{\emph{Natural Language Engineering}} \bibinfo{volume}{7}, \bibinfo{number}{4} (\bibinfo{year}{2001}), \bibinfo{pages}{361--378}.
\newblock


\bibitem[Wallat et~al\mbox{.}(2025)]%
        {wallat2025study}
\bibfield{author}{\bibinfo{person}{Jonas Wallat}, \bibinfo{person}{Abdelrahman Abdallah}, \bibinfo{person}{Adam Jatowt}, {and} \bibinfo{person}{Avishek Anand}.} \bibinfo{year}{2025}\natexlab{}.
\newblock \showarticletitle{A study into investigating temporal robustness of llms}.
\newblock \bibinfo{journal}{\emph{arXiv preprint arXiv:2503.17073}} (\bibinfo{year}{2025}).
\newblock


\bibitem[Yang et~al\mbox{.}(2015)]%
        {yang2015wikiqa}
\bibfield{author}{\bibinfo{person}{Yi Yang}, \bibinfo{person}{Wen-tau Yih}, {and} \bibinfo{person}{Christopher Meek}.} \bibinfo{year}{2015}\natexlab{}.
\newblock \showarticletitle{Wikiqa: A challenge dataset for open-domain question answering}. In \bibinfo{booktitle}{\emph{Proceedings of the 2015 conference on empirical methods in natural language processing}}. \bibinfo{pages}{2013--2018}.
\newblock


\bibitem[Yoon et~al\mbox{.}(2024)]%
        {yoon2024listt5}
\bibfield{author}{\bibinfo{person}{Soyoung Yoon}, \bibinfo{person}{Eunbi Choi}, \bibinfo{person}{Jiyeon Kim}, \bibinfo{person}{Hyeongu Yun}, \bibinfo{person}{Yireun Kim}, {and} \bibinfo{person}{Seung-won Hwang}.} \bibinfo{year}{2024}\natexlab{}.
\newblock \showarticletitle{Listt5: Listwise reranking with fusion-in-decoder improves zero-shot retrieval}.
\newblock \bibinfo{journal}{\emph{arXiv preprint arXiv:2402.15838}} (\bibinfo{year}{2024}).
\newblock


\bibitem[Zheng et~al\mbox{.}(2023)]%
        {zheng2023judging}
\bibfield{author}{\bibinfo{person}{Lianmin Zheng}, \bibinfo{person}{Wei-Lin Chiang}, \bibinfo{person}{Ying Sheng}, \bibinfo{person}{Siyuan Zhuang}, \bibinfo{person}{Zhanghao Wu}, \bibinfo{person}{Yonghao Zhuang}, \bibinfo{person}{Zi Lin}, \bibinfo{person}{Zhuohan Li}, \bibinfo{person}{Dacheng Li}, \bibinfo{person}{Eric Xing}, {et~al\mbox{.}}} \bibinfo{year}{2023}\natexlab{}.
\newblock \showarticletitle{Judging llm-as-a-judge with mt-bench and chatbot arena}.
\newblock \bibinfo{journal}{\emph{Advances in Neural Information Processing Systems}}  \bibinfo{volume}{36} (\bibinfo{year}{2023}), \bibinfo{pages}{46595--46623}.
\newblock


\bibitem[Zhuang et~al\mbox{.}(2022)]%
        {zhuang2022rankt5finetuningt5text}
\bibfield{author}{\bibinfo{person}{Honglei Zhuang}, \bibinfo{person}{Zhen Qin}, \bibinfo{person}{Rolf Jagerman}, \bibinfo{person}{Kai Hui}, \bibinfo{person}{Ji Ma}, \bibinfo{person}{Jing Lu}, \bibinfo{person}{Jianmo Ni}, \bibinfo{person}{Xuanhui Wang}, {and} \bibinfo{person}{Michael Bendersky}.} \bibinfo{year}{2022}\natexlab{}.
\newblock \bibinfo{title}{RankT5: Fine-Tuning T5 for Text Ranking with Ranking Losses}.
\newblock
\newblock
\showeprint[arxiv]{2210.10634}~[cs.IR]
\urldef\tempurl%
\url{https://arxiv.org/abs/2210.10634}
\showURL{%
\tempurl}


\end{thebibliography}

\end{document}